# *LiteCON*: An All-Photonic Neuromorphic Accelerator for Energy-efficient Deep Learning

**DHARANIDHAR DANG,** University of California, San Diego
**BILL LIN,** University of California, San Diego
**DEBASHIS SAHOO,** University of California, San Diego

Deep learning is highly pervasive in today's data-intensive era. In particular, convolutional neural networks (CNNs) are being widely adopted in a variety of fields for superior accuracy. However, computing deep CNNs on traditional CPUs and GPUs brings several performance and energy pitfalls. Several novel approaches based on ASIC, FPGA, and resistive-memory devices have been recently demonstrated with promising results. Most of them target only the inference (testing) phase of deep learning. There have been very limited attempts to design a full-fledged deep learning accelerator capable of both training and inference. It is due to the highly compute and memory intensive nature of the training phase. In this paper, we propose *LiteCON*, a novel analog photonics CNN accelerator. *LiteCON* uses silicon microdisk-based convolution, memristor-based memory, and dense-wavelength-division-multiplexing for energy-efficient and ultrafast deep learning. We evaluate *LiteCON* using a commercial CAD framework (IPKISS) on deep learning benchmark models including LeNet and VGG-Net. Compared to the state-of-the-art, *LiteCON* improves the CNN throughput, energy-efficiency, and computational efficiency by up to 32×, 37×, and 5× respectively with trivial accuracy degradation.

CCS Concepts: • Computer systems organization → Neural networks; Heterogeneous (hybrid) systems; Architectures; • Hardware → Photonic and optical interconnect; Emerging optical and photonic technologies

General Terms: Design, Experimentation, Performance

Additional Key Words and Phrases: Deep learning, on-chip photonics, memristor

## 1. INTRODUCTION

Convolution neural networks (CNNs) have become the go-to solution for a wide range of problems, such as object recognition [1], speech processing and machine translation. Deep CNN models trained with large datasets are highly relevant and critical to ever-growing cloud services, such as face identification (e.g., Apple iPhoto and Google Picasa) and speech recognition (e.g., Apple Siri and Google assistant). However, a CNN algorithm involves a huge volume of computationally intensive convolutions. For example, a basic CNN model created in 2012, AlexNet [2] requires 724M floating point multiply-accumulate (MAC) operations just for inference. The floating-point MAC count is more than 20 times in general for training a CNN. Such increase in the number of MAC makes training three order of magnitude higher compute and memory intensive than inference [5]. As a result, traditional CPUs and GPUs struggle to achieve high processing throughput per Watt [3] for CNN applications. To address this, several FPGA [4] and ASIC [5] approaches have been proposed to accomplish large-scale CNN acceleration.

A CNN comprises two stages: training and inference. Most hardware accelerators for CNNs in prior literature focus only on the inference stage, while the training is done offline using GPUs. However, training a CNN is up to several hundred times more compute and power intensive than its inference [5]. Moreover, for many applications, training is not a one-time activity, especially under changing environmental and system conditions, where re-training of CNN at regular intervals is essential to maintaining prediction accuracy for the application over time. This calls for an energy-efficient training accelerator in addition to the inference accelerator.

Training a CNN in general, employs a backpropagation algorithm which demands high memory locality and compute parallelism. Recently, a few resistive-memory (ReRAM or memristor crossbar) based training accelerators have been demonstrated for CNNs, e.g., ISAAC [5], PipeLayer [6], and RCP [7]. ISAAC and RCP use highly parallel memristor crossbar arrays to address the need for parallel computations in CNNs. In addition, ISAAC uses a very deep pipeline to improve system throughput. However, this is only beneficial when a large number of consecutive images can be fed into the architecture. Unfortunately, during training, in many cases, a limited number of consecutive images need to be processed before weight updates. The deep pipeline in ISAAC also introduces frequent pipeline bubbles. Compared to ISAAC, PipeLayer demonstrates an improved pipeline approach to enhance throughput. However, RCP, DPE, ISAAC, and PipeLayer involve several analog-to-digital (AD) and digital-to-analog (DA) conversions which become a performance bottleneck, in addition to their large power consumption. Also, training in these accelerators involves sequential weight updates from one layer to another. This incurs inter-layer waiting time for synchronization, which reduces overall performance. This calls for an analog accelerator that can drastically reduce the number of AD/DA conversions, and inter-layer waiting time. It has been recently demonstrated that a completely analog matrix-vector multiplication is 100× more efficient than its digital counterpart implemented with an ASIC, FPGA, or GPU [8]. Vandroome et al. in [9] have demonstrated a small-scale efficient recurrent neural network using analog photonic computing. A few efficient on-chip photonic inference accelerators have also been proposed in [10], [11][23]. However, a full-fledged analog deep learning (or CNN to be precise) accelerator that is capable of both training and inference is yet to be demonstrated.

In this paper, we propose *LiteCON*, a novel silicon photonics-based neuromorphic CNN accelerator. It comprises silicon photonic microdisk-based convolution, memristive memory, high-speed photonic waveguides, analog amplifiers. *LiteCON* works completely in the analog domain, therefore, we use the term neuromorphic in it (A neuromorphic system is made up analog components to mimic human brain behavior, in this case CNN, an artificial neural network). The lower footprint, low-power characteristics, and ultrafast nature of silicon microdisk enhance the efficiency of *LiteCON*. *LiteCON* is a first-of-its-kind memristor-integrated silicon photonic CNN accelerator for end-to-end analog training and inference. It is intended to perform highly energy efficient and ultra-fast training for deep learning applications with state-of-the-art prediction accuracy. *The main contributions of this article are summarized as follows:*

- We propose *LiteCON*, a fully analog and scalable silicon photonics-based CNN accelerator for energy-efficient training;

- We introduce a novel compute and energy-efficient silicon microdisk-based convolution and backpropagation architecture;
- We demonstrate a pipelined data distribution approach for high throughput training with *LiteCON*;
- We synthesize the *LiteCON* architecture using a photonic CAD framework (IPKISS [16]). The synthesized *LiteCON* is used to execute four variants of VGG-Net [12] and two variants of LeNet [13], demonstrating up to 30×, 34×, and 4.5× improvements during training, and up to 34×, 40×, and 5.5× during inference, in throughput, energy efficiency, and computational efficiency per watt respectively, compared to the state-of-the-art CNN accelerators.

The rest of the article is organized as follows. Section 2 presents a brief overview of CNNs and prior art. Section 3 provides a gentle introduction of the components used in *LiteCON*. The details of the *LiteCON* architecture are described in Section 4. Section 5 illustrates an example design of *LiteCON* followed by Section 6 which contains the experimental setup, results, and comparative analysis. Lastly, we present concluding remarks in Section 7.

## 2. BACKGROUND AND PRIOR ART

### 2.1 CONVOLUTION NEURAL NETWORKS

CNNs are a class of deep learning network commonly used for analyzing visual imagery for image classification and object detection tasks. A CNN comprises three types of layers: convolution layer (CONV), pooling layer (POOL) and a fully connected layer (FC). Generally, CONV is accompanied with a non-linear activation function, such as ReLU, Tanh, or Sigmoid. CNN operates in two stages: training and inference (testing). In the training phase, the filter weights (and biases) in CONV and FC layers are learnt by using a backpropagation (BP) algorithm. The BP algorithm involves a forward and a backward pass in the deep network. Given a training sample $x$ in the forward pass, the weighted input sum (convolution) $z$ is computed for neurons in each layer $l$ with some initial filter weights $w$ (and bias $b$) followed by neural activation $\sigma(z)$ (ReLU($z$) in our work), and POOL. The final layer $L$ computes the output label of the overall network for every forward pass. This can be summarized as follows:

**Forward Pass:** For each layer $l$,

$$z^{x,l} \leftarrow w^l a^{x,l-1} + b^l \tag{1}$$

$$a^{x,l} \leftarrow \sigma(z^{x,l}) \tag{2}$$

The output error in the final prediction $\delta^{x,L}$ is a result of errors induced by the neurons in each hidden layer during the forward pass. To determine the error contribution of a neuron in the previous layer i.e., $\delta^{x,l}$, the final error is back propagated through the network starting from the output layer. This can be summarized as follows:

**Output error:** At the final layer $L$,

$$\delta^{x,L} \leftarrow \nabla_a C_x \odot \sigma'(z^{x,L}) \tag{3}$$

**Backward Pass:** For each layer $l$,

$$\delta^{x,l} \leftarrow ((w^{l+1})^T \times \delta^{x,l+1}) \odot \sigma'(z^{x,l}) \quad (4)$$

Here, $\nabla_a$ is gradient of $a^{x,l}$, and $\sigma'(z^{x,L})$ is derivative of $\sigma(z^{x,L})$. These error contributions are necessary to update the filter weights $w$ and biases $b$ in the respective layers using a gradient descent method. In gradient descent, the forward and backward pass happen iteratively until the cost function is minimized and the network is trained. This can be summarized as follows:

**Gradient Descent:** For each layer $l$ and $m$ training samples with learning rate $\eta$,

$$w^l \leftarrow w^l - \frac{\eta}{m}\sum_x \delta^{x,l} \times (a^{x,l-1})^T \qquad (5)$$

$$b^l \leftarrow b^l - \frac{\eta}{m}\sum_x \delta^{x,l} \qquad (6)$$

The next section presents the details of the proposed *LiteCON* architecture.

## 2.2 PRIOR ART

To achieve high speed and energy efficient deep learning, researchers recently have demonstrated photonic accelerators by deploying microring weight banks [12][21], Mach-Zehnder interferometers [10][22] and multilayer diffractive optical elements [13]. Compared to these optical devices, silicon photonic microdisk (MD) has a smaller chip-area, ultrafast nature, and lower power characteristics [14]. Moreover, most silicon photonic accelerators cannot achieve the state-of-the-art inference accuracy even for small datasets. For example, none of the photonic accelerators perform with an accuracy > 97% on small MNIST dataset, while recent CNNs easily reach > 99.77% [2] accuracy on the same dataset. This is due to the large noises accumulated during fully-optical inferences. Our proposed design based on MD addresses these bottlenecks.

## 3. COMPONENTS OVERVIEW

CMOS compatible components such as photonic waveguides, silicon microdisks (MDs), photodiodes and multi-wavelength LED array are used for on-chip photonic signaling [15]. An MD is a circular shaped photonic structure which is used to modulate electronic signals into a photonic signal at the transmission source in a waveguide. MDs are also used to couple or filter out light from the waveguide at the destination. Each MD modulates light of a specific wavelength and its geometry (radius to be precise) determines its wavelength selectivity. We can also inject (or remove) charge carriers to (from) an MD or heat it to alter its operating wavelength.

In a typical high-bandwidth photonic link, an LED array (either on the board or on a 2.5D interposer) generates multiple wavelengths, which are coupled by an optical grating coupler to an on-chip photonic waveguide. The phenomenon of using multiple wavelengths to transmit many streams of bits simultaneously is referred to as dense-wavelength-division-multiplexing (DWDM). To enable processing of these photonic signals, the on-chip photonic waveguide propagates the input optical power to the destination where they are captured by photodiodes and are converted to electronic data. These components are the building blocks of the proposed *LiteCON* architecture.

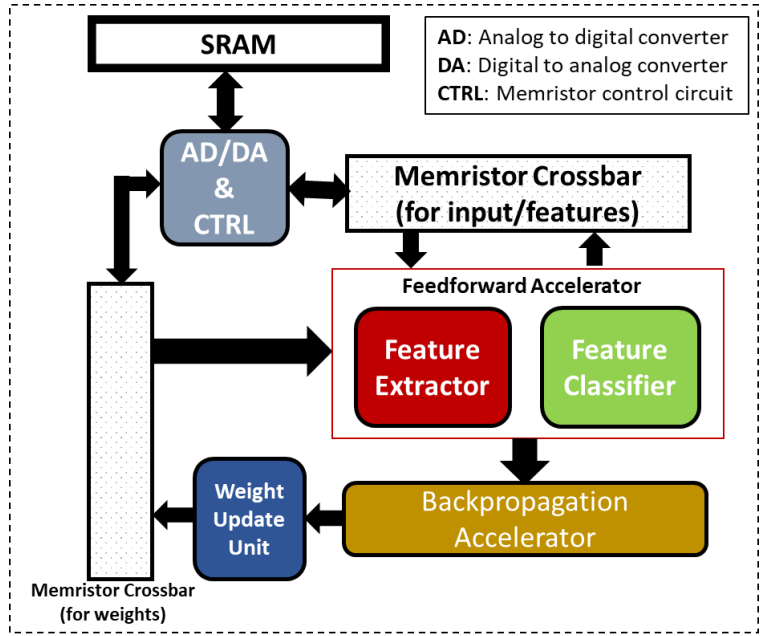

Figure 1: An overview of *LiteCON* architecture.

## 4. *LiteCON* ARCHITECTURE

**Overview**: Our proposed *LiteCON* architecture is a fully analog, scalable silicon photonics-based CNN accelerator design. Unlike previously proposed CNN accelerators [4-6], *LiteCON* accelerator enables fully analog end-to-end training and inference for CNN. Fig. 1 gives a high-level overview of the *LiteCON* architecture. As shown in the figure, *LiteCON* comprises four major parts: feature extractor, feature classifier, backpropagation accelerator, and weight update unit. The feature extractor (FE) and the feature classifier (FC) are made up of multiple silicon microdisk-based convolution layers, operational amplifier (OPAMP)-based ReLU layers, and pooling layers. *Together, FE and FC make the feedforward CNN accelerator*. *The backpropagation accelerator is built using silicon microdisks, splitters, and multiplexers*. And, *LiteCON*'s weight update unit is designed by deploying a group of memristors.

### 4.1 FEEDFOWARD ACCELERATOR

In this paper, we consider image dataset as input and its classification as the task to be performed by *LiteCON*. The digital input data is stored in SRAM. The feedforward accelerator in *LiteCON* architecture (see Fig. 1) performs feedforward feature extraction (FE) followed by feature classification of input images. It operates in four stages: (a) data reading, (b) feature extraction, (c) feature classification, and (d) data writeback. The details are as follows.

*4.1.1 Data Reading*: *LiteCON* is designed to convolve an input of 28×28 pixels at a time, i.e., one *LiteCON* cycle. Therefore, it requires 64 *LiteCON* cycles to execute a 224×224 image (typical size of an ImageNet image). Please note that a *LiteCON* cycle is different from its clock cycle. Here, one *LiteCON* cycle refers to the complete feature extraction and feature classification of a 28×28 image. The SRAM in *LiteCON* is of size 256 KB (dual data rate, 64 bits) to store the five images of size $224{\times}224$. In a pipelined fashion, four blocks of 28x28 pixels are written into the memristor crossbar (capable of storing four 28x28 pixels) via an n-channel DAC and memristor controller (ref: Fig.2). The crossbar can be understood as a high-speed cache for *LiteCON*.

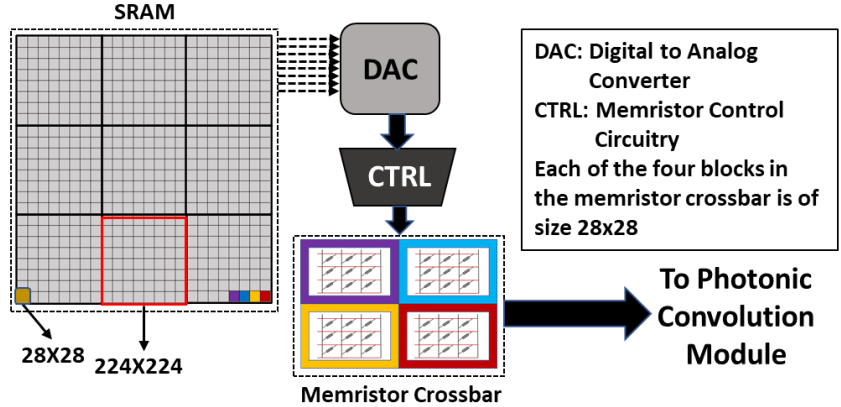

Figure 2: Data Reading from SRAM to memristor crossbar

***4.1.2 Feature Extraction:*** The feature extraction (FE) in our architecture is carried out using multiple FE stages ($FE_i$). Each FE stage comprises multiple photonic convolution layers (PConv), an analog amplifier (OPAMP)-based ReLU layer, another OPAMP-based pooling (POOL) layer, and an interface layer. *LiteCON's* FE adopts a completely analog computing paradigm by avoiding inter-layer A-to-D (Analog-to-Digital) and D-to-A (Digital-to Analog) conversions compared to state-of-the-art CNN accelerators [5], [6] which use analog memristive convolution and digital CPU/GPU based ReLU and pooling.

***Photonic Convolution (PConv)***:

PConv is the first layer of an FE. The PConv is based on the principle of analog multiplication using silicon microdisk [14]. *A silicon microdisk is used for analog amplitude modulation of a light carrier. In its simplest term, analog amplitude modulation is the multiplication of a scalar input with an analog signal. The authors in [14] have demonstrated photonic modulator based analog multipliers*. In our design, a PConv is made to convolve 28×28 pixels at a time. It can be scaled up depending upon the requirements. A PConv comprises (i) an array of leds capable of generating up to $N$ wavelength carriers; (ii) a DWDM multiplexer, splitter, and waveguide arrangement to accommodate all the carriers into one channel; (iii) $N{\times}(M{+}1)$ number of microdisks ($N{\times}M$ for microdisk multiplications and another $N$ for weight modulation); and (iv) $N$ photodiode.

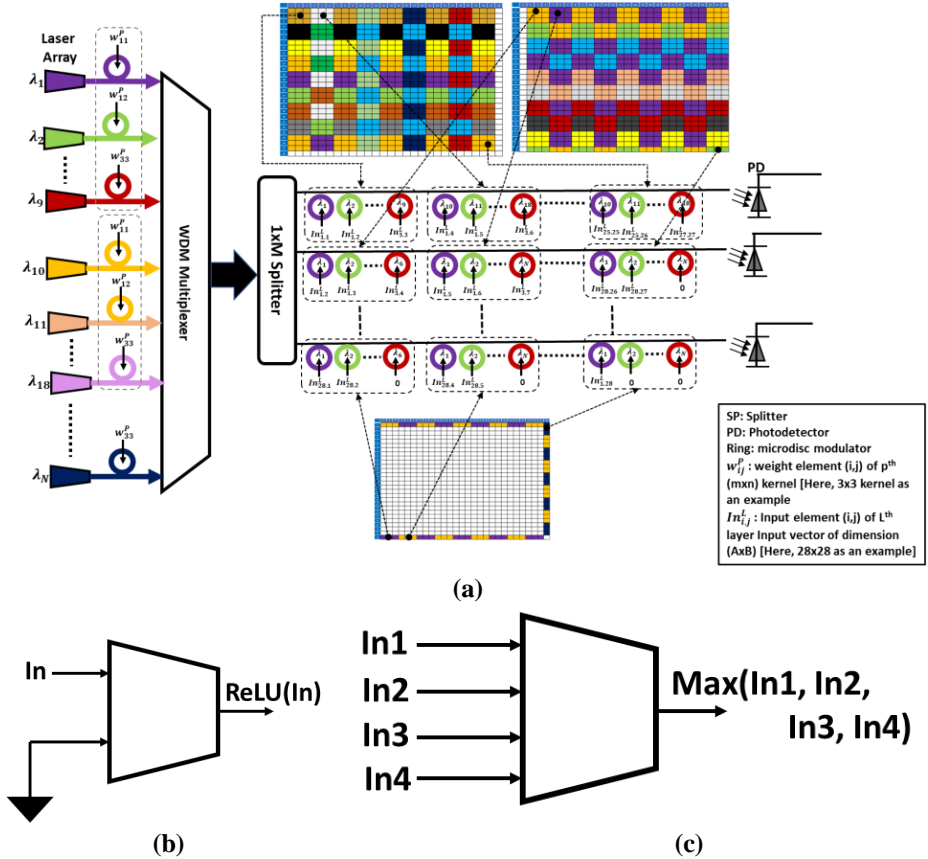

**(a)**

**(b)** **(c)**

Figure 3: (a) Logical microarchitecture of MD-based photonic convolution; (b) OPAMP-based ReLU layer; and (c) four-input OPAMP-based POOL layer.

Convolution in deep learning operates with kernels (or filters) of several sizes such as 1×1, 2×2, 3×3, 4×4, and so on. The widely adopted models that we consider in this paper (refer: Table I) comprise 3×3 filters. Hence, Fig.3a depicts photonic convolution based on a 3×3 filter. To start with, each of the $N$ wavelength channels from the LED array is integrated with a microdisk of respective wavelength. All microdisks are divided into $K$ groups ($K \times 9 = N$), each having 9 microdisks. All these groups of microdisks are then modulated with weight values ($w_{11}^L, w_{12}^L, ... w_{33}^L$) stored in the memristor crossbar (one part of memristor crossbar stores weights and another part stores input data or features obtained in hidden layers). Here $w_{ij}^L$ is the weight (i, j) of a filter in the $L^{th}$ layer of convolution. All the $N$ modulated wavelengths are multiplexed into one waveguide by a DWDM multiplexer following which the multiplexed light is split into $P$ equal channels each carrying all the modulated wavelengths. Each channel is equipped with 784 microdisks (it can be scaled up or down depending upon the input size, here $28 \times 28$). Now, in each channel, pixel values stored in the memristor crossbar are modulated into individual wavelength by the microdisks. As shown in Fig. 3a, the first group of 9 pixel (a matrix) is modulated by the first group of 9 microdisks. The pixels for a channel are chosen in such a way that there is no same pixel modulated to two wavelengths in the same channel (to avoid data collision). That way, in each wavelength carrier, using the multiplication principle of a microdisk, an input pixel $In_{xy}$ is multiplied by a weight value $w_{ij}^L$. Please note that convolution at its core is nothing but a sum of input and wight multiplication. Finally, the multiplexed light from each channel is captured by an array of photodiode. Each photodiode is designed to capture nine consecutive wavelengths. For example, the first photodiode integrated with the first channel captures ($w_{11}^L \times In_{11} + w_{12}^L \times In_{12} \dots + w_{33}^L \times In_{33}$) which is nothing but the 1$^{st}$ convolved matrix. Similarly, other convolved matrices are captured.

Example of a simple PConv:

Let us assume there are 9 pixels in an input. The nine pixels are stored as analog input in the memristor crossbar as $In_{11}, In_{12}, \dots In_{33}$. For simplicity of understanding, suppose there are 9 weights or a 3×3 Filter. They are $w_{11}, w_{12}, \dots w_{33}$. The weights are modulated onto 9 wavelength channels and then passed in a single multiplexed waveguide. Now, each input pixel $In_{xy}$ is modulated by a microdisk into a weight-carrying channel. As in, the violet color microdisk modulates $In_{11}$ into the channel carrying weight $w_{11}$. Thus, microdisk performs amplitude modulation or in other words multiplication; so that channel now carries $w_{11} \times In_{11}$. Similarly, other channels end up carrying $w_{12} \times In_{12}, \dots w_{33} \times In_{33}$. At the end, all these photonic signals are captured together by a photodiode as sum total, i.e., $w_{11} \times In_{11} + w_{12} \times In_{12} + \dots w_{33} \times In_{33}$. That is how photonic convolution works.

***Electronic ReLU and Pooling:***

Neural activation in CNN can be performed by a variety of non-linear functions such as Sigmoid, Tanh, ReLU (rectified linear unit), and etc. ReLU is widely used for its simplicity of implementation and exemplary performance. Therefore, we consider a ReLU-based neural activation circuit. The following equation explains the working of a ReLU unit.

$$ReLU\ (z) = z\ if\ z > 0$$
$$= 0\ if\ z \leq 0 \qquad (7)$$

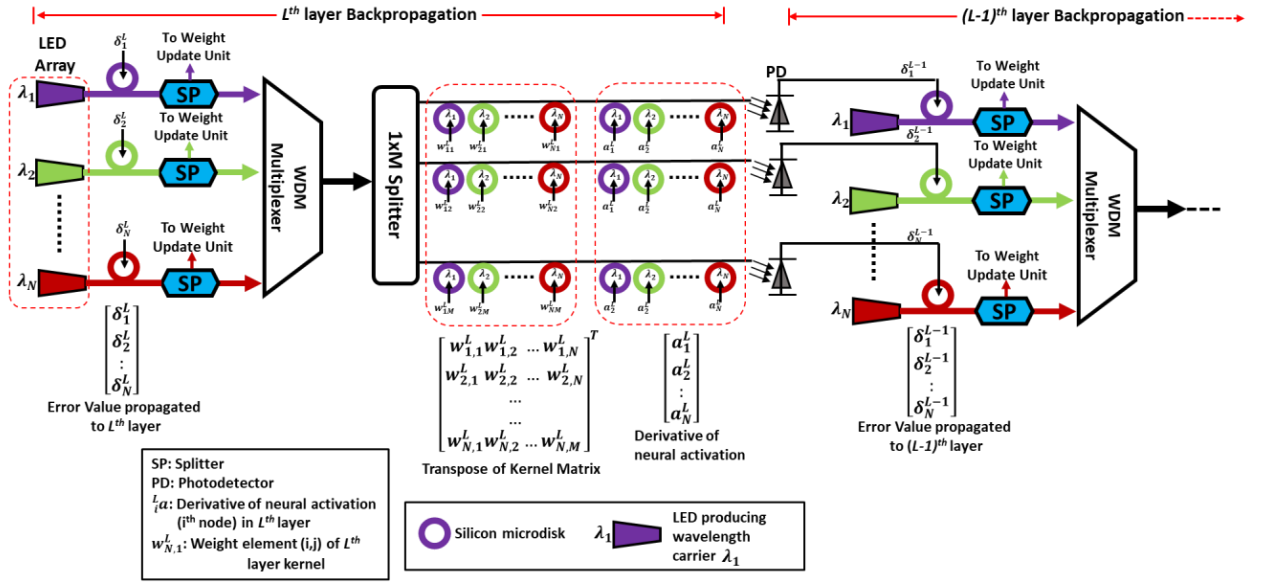

Figure 4: (a) Schematic view of *LiteCON*'s backpropagation accelerator

We deploy an operational amplifier (OPAMP) to mimic such a function as shown in Fig. 3b. Because Eq.7 can be seen as an example of a comparator. And analog OPAMP does the same thing. It takes two inputs and generates a target output based on the comparison. The output of photodiode from Fig.3a is fed as input to the OPAMP for ReLU operation. Please note that the OPAMP circuitry can be reconfigured to mimic other neural activation functions. The details of OPAMP mimicking other functions are omitted due to brevity.

The next operation in FE is pooling which is used to reduce the feature size and keeps spatial invariance. It does so by taking the average or maximum of multiple elements of a feature vector. We choose maximum for its superior accuracy in a variety of applications. Pooling is also a comparator function at a fundamental level. Four or nine outputs (reason: 2×2 or 3×3 is the typical pooling size) from ReLU units are fed as input to an OPAMP-based comparator which then selects the maximum value as the spatially-invariant pooling output (ref: Fig.3c). *The outputs from all the comparator are the extracted feature which is stored back in the memristor crossbar for the next layer of FE*. When features go through all the FE stages, obtained features are stored back in SRAM.

***4.1.3 Photonic Feature Classification:*** After feature extraction is performed using the FE stages (by PConv, ReLU, and pooling), features are brought back from the SRAM via the memristor crossbar to undergo feature classification phase. In CNN, the feature classification segment can be seen as a special case of convolution, where each extracted feature map uses the largest possible kernel. In other words, feature classification comprises one or more fully-connected (FC) layers (In FC, each element from one layer is connected to all the elements in the next layer).

*LiteCON* employs microdisk-based matrix vector multiplier (M-MVM) to implement the FC layer identical to PConv (refer Section 4.1.2). In FC, each wavelength channel is modulated with a different weight value (unlike 9 weights in group as in the case of PConv) to ensure fully-connected network. When all the features from the feature extraction stage are brought back from the SRAM and available in the memristor crossbar, the features are fed to the FC layer. As an example, we consider 512 features coming from the feature extraction (FE) stages. VGG and LeNet operate on a 7×7 kernel in FC. Therefore, each feature is a 7×7 matrix. Therefore, 49 wavelength carriers from an LED array are modulated with 49 weights by microdisks. After multiplexing and splitting into 512 equal channels (similar to PConv), each channel is matrix multiplied with one feature. The obtained output at the photodiode is fed to ReLU followed by pooling (if required). Then the results are fed to the next FC layer (if present in the model). *After features go through all the FC layers, we obtain the classified output*. During training, the classified outputs from the final FC and target outputs are input to an analog subtraction unit, the result (or error vector) of which is fed to the backpropagation architecture, as discussed next.

## 4.2 BACKPROPAGATION ACCELERATOR

*LiteCON*'s backpropagation (BP) accelerator employs silicon microdisks, photodiodes, multiplexers, and splitters to perform completely analog matrix-multiplication and other arithmetic operations, similar to the PConv. In contrast, previously proposed CNN accelerators [5], [6] adopt a hybrid approach by using analog memristors for matrix multiplications and digital CPU/GPU for other arithmetic operations, which requires performance hindering A-to-D and D-to-A conversions.

Fig. 4a illustrates the microarchitecture of the proposed BP accelerator design. It is based on photonic matrix-vector multiplication using silicon microdisks (MDs). We use MDs for their smaller footprint, high accuracy and quality factor, and low-power nature. We now explain the operation of the proposed BP architecture. As discussed in Eq. (3), the error at the final layer ($l=L$) of BP is

$$\delta^{x,L} \leftarrow \nabla_a C_x \odot \sigma'(z^{x,L}).$$ Here, $\nabla_a C_x$ is rate of change in output w.r.t the output activation

(i.e., difference between actual classified output from feedforward accelerator and the target output stored in memristor crossbar). $\sigma'(z^{x,L})$ is the derivative of the ReLU layer in the final FC stage of the CNN architecture. Outputs from the final FC stage of the CNN architecture are fed to an analog subtraction and multiplication unit (microdisk multiplier) to determine $\delta^{x,L}$. Applying Eq. (4) and using the computed $\delta^{x,L}$, we calculate error for the $(L\text{-}1)^{th}$ layer as below:

$$\delta^{x,L-1} \leftarrow ((w^L)^T \times \delta^{x,L}) \odot \sigma'(z^{x,L-1}) \qquad (8)$$

where, $w^L$ is weight matrix (stored in memristor crossbar) obtained from $L^{th}$ layer of feedforward CNN architecture. Fig.4a shows the backpropagation between the final layer $l=L$ and its penultimate layer $l=L\text{-}1$. As illustrated, there are N number of wavelength carriers coming from an LED array. The value of N for a layer equal to the output feature size for the corresponding layer in the feedforward accelerator, e.g., N equals 49 (7×7) for the last layer. Each wavelength in layer $L$ is modulated with error $\delta^{x,L}$ by a MD tuned to that wavelength. In Fig.4a, the violet MD is tuned to modulate $\lambda_1$. Let us suppose the $j^{th}$ MD's output is $MD_j = \delta_j^{x,L} * A \sin(\frac{2\pi}{\lambda_j} t + \emptyset)$ ($A \sin(\frac{2\pi}{\lambda_j} t + \emptyset)$ represents the photonic carrier with wavelength $\lambda_j$ and phase difference $\emptyset$). Each $MD_j$ is split into two equal parts. The first part is sent to the weight-update circuitry (explained at the end of this section) to update the corresponding weights in the feedforward accelerator. The other part is fed to a DWDM multiplexer. A DWDM multiplexer is used to combine multiple light wavelengths into a single multi-wavelength carrier. After multiplexing, the multiplexed photonic data is split into $M$ parts by an optical splitter where $M$ equals the number of neurons in layer $L\text{-}1$. Each part is fed to a multi-wavelength waveguide. As a result, in each waveguide there are N wavelengths each carrying data $\delta_{j,n}^{x,L} * B \sin(\frac{2\pi}{\lambda_j} t + \emptyset)$, where $1 \leq n \leq N, B = \frac{A}{2N}$. Each weight $w_{ij}^L$ of the transpose of $w^L$ obtained from the memristor crossbar is modulated by an MD to a light carrier. This results in:

$$D_{i,n} = w_{ij}^L * \delta_{j,n}^{x,L} * A \sin(\frac{2\pi}{\lambda_j} t + \emptyset) \qquad (9)$$

Now, each $D_{i,n}$ is modulated with $a_n^L$ which is a derivative of the ReLU functions of layer $L\text{-}1$ (equal to $\sigma'(z^{x,L-1})$ in Eq. (8)). Then, $D_{i,n}$ becomes,

$$D_{i,n} = w_{ij}^L * \delta_{j,n}^{x,L} * a_n^L * A \sin(\frac{2\pi}{\lambda_j} t + \emptyset) \qquad (10)$$

Next, a photodiode is used to demodulate photonic data from each waveguide. The photodiode captures the combined output $D_{i,n}$ for all wavelengths in a waveguide which is nothing but the matrix-vector multiplied error vector identical to Eq. (8). The output of each photodiode is passed through a signal conditioning circuit to remove unwanted noises. Details of the conditioning circuit are omitted for brevity. The output from the signal conditioning circuit looks as follows:

$$\delta^{x,L-1} = ((w^L)^T \times \delta^{x,L}) \odot a^L \qquad (11)$$

where, $\delta^{x,L-1}$ is the error to be propagated from layer *(L-1) to (L-2)*. The above procedure is continued until the 1$^{st}$ layer of *LiteCON* is reached. While doing the backpropagation, the error value in each layer is also sent to the corresponding weight-update circuit, which is discussed in more detail below.

**Weight-update circuitry:**

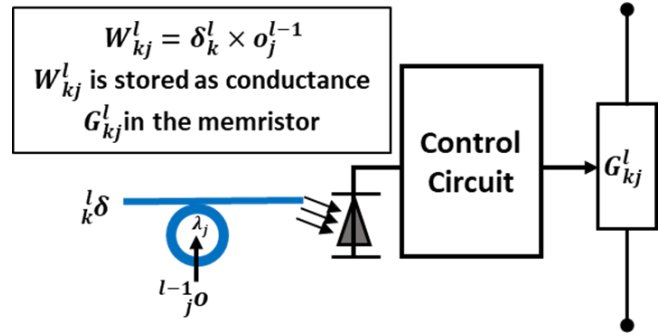

Figure 5: Weight-update circuitry for any layer *l*

For weight update, each element of a weight kernel in any layer *l* of the CNN architecture can be written as $w_{k,j}^l$. Please note that *l=L* for the final layer. Each $w_{k,j}^l$ is stored in a memristor cell of a memristor crossbar in layer *l* as $C_{k,j}^l$ (which is the conductance of a memristor cell). The weight-update equation for $w_{k,j}^l$ (or, $C_{k,j}^l$) can be written as per Eq. (5), as follows:

$$C_{new(k,j)}^l \leftarrow C_{old(k,j)}^l - \frac{\eta}{m} \times \delta_k^l \times O_j^{l-1} \qquad (12)$$

where, $O_j^{l-1}$ is the $j^{th}$ output from the POOL of the *(l-1)* layer of the CNN architecture. Fig.5 illustrates the weight-update circuitry for any layer *l*. As shown in Fig.4b, $\delta_k^l$ is obtained from the BP architecture as a photonic signal. $O_j^{l-1}$, which is collected from the memristor crossbar (for data storage), is used to modulate the light carrier carrying the error value $\delta_k^l$. The modulated output is demodulated using a photodiode and then sent to a signal conditioning circuit. In the signal conditioning circuit, first the analog signal is filtered (from noises) and passed through a subtractor to obtain new $C_{k,j}^l$ as depicted in Eq. 12). The previous conductance or weight value $C_{old(k,j)}^l$ is fed to the subtractor circuit from the $l^{th}$ layer BP architecture. The new conductance value $C_{k,j}^l$ is now fed to the equivalent memristor control circuit to update its weight value. The conditioning circuit, as well as the memristor control circuit, is inspired from [7].

## 5. *LITECON* CASE STUDY

In this section, we present the working of the proposed pipelined *LiteCON* architecture for a CNN benchmark VGG [16] on the ImageNet dataset [17]. In our experiments, we consider all variants of the VGG [16] and LeNet [18] benchmarks as shown in Table I. We integrate the PConv layer, ReLU layer, POOL layer, and FC layer based on VGG-A model for this case study as shown in Fig. 6a. See that for one convolution in FE$_1$ for VGG-1 (ref: Table I), there is an equivalent PConv in FE$_1$ of Fig. 6a; similarly, for two back-to-back convolutions in FE$_3$, there are two back-to-back PConv in FE$_3$ of

Fig. 6a. The backpropagation accelerator is connected to the feedforward accelerator as follows: BP-1 with FE$_1$, BP-2 with FE$_2$, BP-3 with FE$_3$, and so on. The rest of the section discusses how *LiteCON* mimics VGG-A.

VGG for the ImageNet dataset operates on a 224×224 image input. As mentioned earlier, *LiteCON* is designed to convolve 28×28 pixels at a time, i.e., one *LiteCON* cycle. Therefore, it requires 64 *LiteCON* cycles to execute a 224×224 image. The SRAM register array in *LiteCON* is of size 256 KB to store five images of size 224×224. PConv performs feature extraction on a 28×28 input data at a time in a pipelined manner.

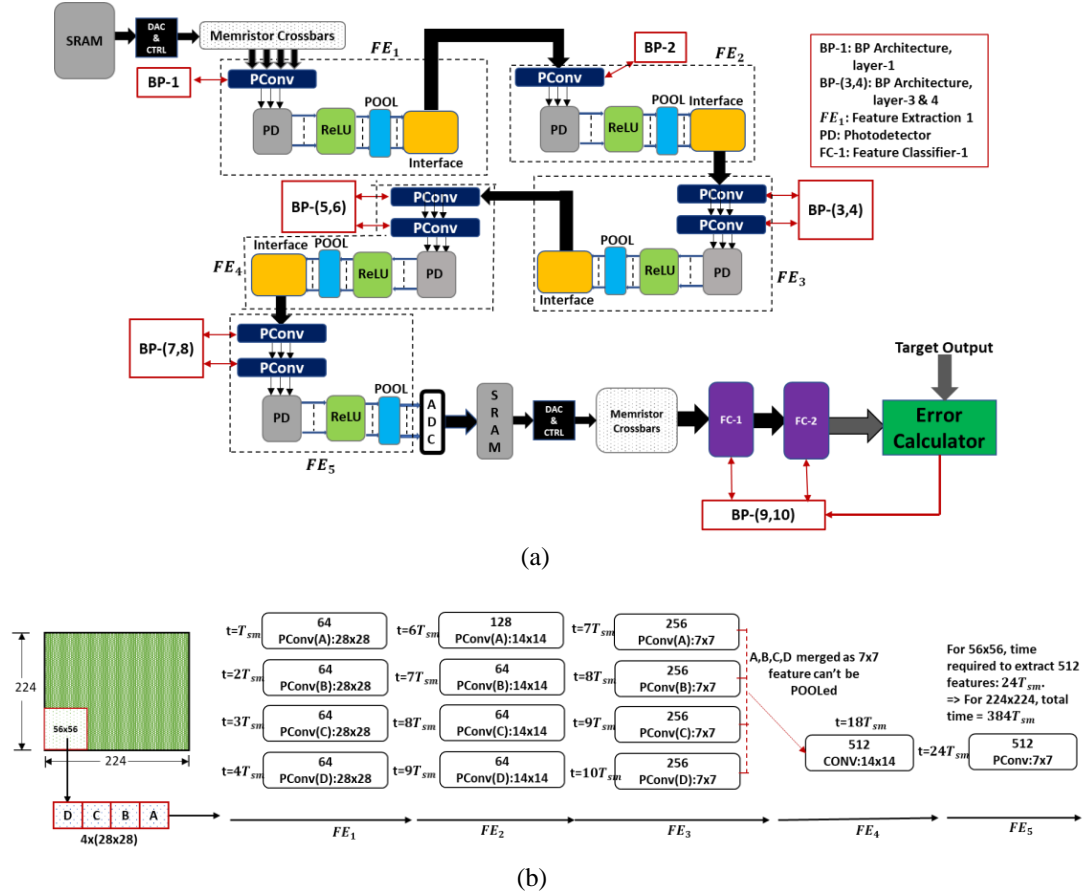

(a)

(b)

Figure 6: (a) VGG-A implemented on *LiteCON* (b) Pipelined dataflow in feedforward operation in *LiteCON*.

Fig. 6b demonstrates the pipelined dataflow of the feedforward operation in *LiteCON*. We consider a 2.5 GHz clock. Therefore, the clock cycle period T$_{sm}$ = 400 ps. As shown in Fig. 6b, at t=T$_{sm}$, the first set of 28×28 pixels from SRAM (i.e., A) are convolved (64 filters/features) and are stored in memristor crossbars (for data storage) (the yellow interface module in Fig.5a represents data transfer into memristors in the peripheral circuit). To illustrate the pipelined approach, we explain the convolution of another three set of 28×28 pixels namely, B, C, and D. Note that PConv convolves a 28×28 input in one clock cycle (ref: Section 4.1.2). As FE$_1$ for VGG-A consists of one convolution layer (see Table I), convolved photonic outputs of PConv-1 of $FE_1$ is sent to the ReLU layer through the photodiode followed by the POOL layer. The time required for convolved data of one FE to arrive at the next FE, $T_{FE}$ = photodiode conversion time + ReLU time + POOL time + interface time = 20 ps + 10 ps + 10 ps + 10 ps = 50 ps. From $t=T_{sm}$ to $t=2T_{sm}$, PConv(A) outputs from the peripheral circuit of $FE_1$ are photodiode-converted, ReLUed and POOL'ed, and then fed to $FE_2$.

TABLE I. CNN BENCHMARK CONFIGURATION FOR VGG, LeNeT [Read (i×j, m, k) as filter size: i×j, number of such filters: m, and number of back-to-back convolutions in a layer: k]

| | $FE_1$ | $FE_2$ | $FE_3$ | $FE_4$ | $FE_5$ | |
|---|---|---|---|---|---|---|
| **VGG-A** | 3×3, 64, 1 | 3×3, 128, 1 | 3×3, 256, 2 | 3×3, 512, 2 | 3×3, 512, 2 | FC-4096,2 FC-1000, 1 |
| **VGG-B** | 3×3, 64, 2 | 3×3, 128, 2 | 3×3, 256, 2 1×1, 256, 1 | 3×3, 512, 2 1×1, 256, 1 | 3×3, 512, 2 1×1, 256, 1 | |
| **VGG-C** | 3×3, 64, 2 | 3×3, 128, 2 | 3×3, 256, 3 | 3×3, 512, 3 | 3×3, 512, 3 | |
| **VGG-D** | 3×3, 64, 2 | 3×3, 128, 2 | 3×3, 256, 4 | 3×3, 512, 4 | 3×3, 512, 4 | |
| **LeNET-A** | 3×3, 6, 1 | 3×3, 6, 1 | 3×3, 16, 2 | 3×3, 16, 4 | 3×3, 120, 1 | FC84,1 |
| **LeNET-B** | 3×3, 6, 1 | 3×3, 6, 1 | 3×3, 256, 1 | 3×3, 16,6 | 3×3, 120, 1 | |

There can be 8 such data movements as $\frac{T_{sm}}{T_{FE}}=8$. In one data movement, 4 28×28 features can be processed. Therefore, at $t=2T_{sm}$, 32 PConv(A) features arrive at $FE_2$. Similar to PConv(A), from $t=2T_{sm}$ to $t=3T_{sm}$, 32 PConv(B) features; from $t=3T_{sm}$ to $t=4T_{sm}$, 32 PConv(C) features; from $t=4T_{sm}$ to $t=5T_{sm}$, 32 PConv(D) features are convolved and stored in the peripheral circuit of $FE_2$. After this, from $t=5T_{sm}$ to $t=6T_{sm}$, the remaining 32 PConv(A) features in $FE_1$ are convolved in $FE_2$. In this way, by $t=6T_{sm}$, all the 64 PConv(A) features in $FE_1$ are convolved with 128 $FE_2$ filters to produce 128 features and stored in the memristors of its peripheral circuit. Similarly, remaining 32 B, C, and D features are convolved and stored (Fig. 6b) by $t=7T_{sm}$, $t=8T_{sm}$, and $t=9T_{sm}$ respectively. $FE_1$ has 64 features, $FE_2$ has 128 features, $FE_3$ has 256 features, etc, as per the VGG-A configuration (Table I). It is important to note that 64 PConv(A) features from $FE_1$ are convolved with 128 kernels/filters to produce 128 PConv(A) features for $FE_2$. Similarly, 128 PConv(A) features from $FE_2$ are convolved with 256 kernels to produce 256 PConv(A) features for $FE_3$.

A, B, C, and D are convolved separately until $t = 10T_{sm}$ when all of them arrive at $FE_3$ as 256 7×7 features each. Now, all of these features are merged together to form 256 28×28 features. Therefore, it will require another $8T_{sm}$ time (i.e., $t=10T_{sm}$ to $t=18T_{sm}$) to send 256 28×28 features from $FE_3$ and convolve them as 512 14×14 features at $FE_4$. Similarly, convolution, ReLU, and POOL are performed in $FE_4$ and $FE_5$. As illustrated in Fig. 6b, at $t=24T_{sm}$, 512 features are obtained from $FE_5$ for 56×56 pixels. As shown in Fig. 5a, features from $FE_5$ are stored in SRAM until all the 224×224 pixels are extracted. For 224×224 pixels, it will take 16×24$T_{sm}$=384$T_{sm}$=153.6ns (in 24$T_{sm}$, 4 28×28 pixels are convolved; therefore 16×24$T_{sm}$ for 224×224 pixels). After this, all the features are retrieved from SRAM and fed to FC for feature classification. The first FC operation requires ($T_{sm} + T_{FE}$) time as FC is identical to FE. The second FC operation requires $T_{FE}$ time as no more SRAM or memristor read is needed. This means that *LiteCON* requires 153.6 ns (for FE) +$T_{sm}$ + 2$T$ = 154 ns, for one forward pass. After a forward pass, the FC output is sent to the BP architecture for backpropagation. Each layer in BP requires $T_b$ units of time where $T_b$ = (error modulation to light carrier) + (split time) + (WDM multiplexing time) + (split time) + (weight modulation time) + (ReLU function derivative modulation time) + (photodiode time) = 10 ps + 10 ps + 10 ps + 10 ps + 10 ps + 10 ps +20 ps = 80 ps. It takes $6T_b$ units of time to complete one backward pass. *In summary, LiteCON requires 154 ns for one forward pass and 80 ps for a backward pass. The ultra-fast nature of photonic interconnects allows for high-speed backpropagation in LiteCON.*

## 6. EXPERIMENTAL ANALYSES

### 6.1 DESIGN METHOLOGY

We use IPKISS [19], a commercial photonic CAD toolchain, to design and synthesize all of the photonic components in *LiteCON*. The synthesized components are integrated together to build *LiteCON*. For all of the photonics components, we consider a 32nm IPKISS library. We developed a C++ based architectural simulator which takes device- and link-level parameters from IPKISS, to estimate performance of *LiteCON* accelerator for several benchmarks.

***6.1.1 Power, Area, and Performance Models:*** We use Caphe [19] for modeling power and area of all photonic elements such as microdisks, DWDM multiplexers, waveguides, leds, etc. The energy, timing, and area parameters for memristor crossbars are obtained from [6]. For DAC, we deploy an integration and fire mechanism identical to PipeLayer [6] in our design. The power, latency, and area models are adapted accordingly from PipeLayer. We also obtain power, timing, and area parameters of the ADC from [5], used in the FC layer of *LiteCON*. All these parameters are listed in Table II.

We use TensorFlow [20], a widely-used deep learning framework, to train the datasets in conjunction with photonic component results from IPKISS. We manually map each of our benchmarks in waveguides, ReLU, max-pool, and FC of *LiteCON*. This ensures zero pipeline hazards between any two layers in *LiteCON*. We compare the performance of *LiteCON* with a state-of-the-art CNN accelerator, namely PipeLayer [6] and the latest GPU (obtained from [6]).

For comparison, we evaluate the following metrics: *Throughput* is the total number of operations per unit time (GOPS/s); *Computational efficiency per Watt* represents throughput per unit area per Watt (GOPS/s/Watt/mm$^2$); *Energy efficiency* refers to the number of fixed-point operations performed per watt (GOPS/s/W); and lastly, *Prediction error rate* is the percentage of error in inferring any datasets. Please note that all the results in our analysis are based on an 8-bit weight resolution as the ADC/DAC are of 8-bit resolution.

***6.1.2 Benchmarks and Datasets:*** We execute two widely used CNN benchmarks: VGG-Net [16] and LeNet [18] in *LiteCON*. We consider four variants of the VGG benchmark: VGG-A, VGG-B, VGG-C, and VGG-D and two variations of LeNet (LeNet-A and LeNet-B) as depicted in Table I. For a fair comparison, the configuration of all stages of VGG-Net and LeNet benchmarks identical to [6]. For VGG, we use ImageNet dataset [17] having 224×224 images. We consider a subset of ImageNet, i.e., 1M images with 1000 labels. For LeNet, we use 60,000 28×28 images of MNIST datasets [18] for training and 10,000 28×28 images for testing, with 10 labels.

TABLE II. PARAMETRIC DETAILS

| Components | Parameters | Values | Power (mW) | Area (mm²) |
|---|---|---|---|---|
| SRAM register | Size | 2KB | 10 | 0.2 |
| | Count | 128 | | |
| DAC | Resolution | 8-bit | 4.374 | 0.000208 |
| | Frequency | 1.2 Gbps | | |
| | Channel | 64 | | |
| | Count | 208 | | |
| ADC | Resolution | 8-bit | 490 | 0.294 |
| | Frequency | 1.2 Gbps | | |
| | Count | 245 | | |
| Memristor Crossbar (for weights and data) | Size | 64 KB | 30 | 0.5 |
| Microdisk | Time | 20ps | 1080.8 | 39.38 |
| | Count | 62720 | | |
| Photodiode | Time | 20ps | 1080.8 | 39.38 |
| | Count | 62720 | | |
| Trans-Impedance-Amplifiers (TIA) | Time | 10ps | 0.18 pJ/bit | 0.28 |
| | Count | 62720 | | |
| WDM coupler | Count | 16 | 0 | 0.00028 |
| WDM de coupler | Count | 16 | 0 | 0.00028 |
| OPAMP | Time | 20 ps | 0.05 | 0.0045 |
| | Count | 980 | | |
| LED | Wavelengths | 16 | 32000 | 0.384 |
| | Count | 6 | | |
| Waveguide | DWDM | 16 | 0 | 80 |
| | Width | 450 nm | | |
| | Count | 520 | | |

## 6.2 PERFORMANCE ANALYSIS

Fig.7a and 7b present throughput of the proposed *LiteCON* and PipeLayer [6] compared to the baseline GPU implementation results, also from [6], during training and inference respectively. The GPU-based accelerator performs with an average training throughput of 306 GOPS/s and an average inference throughput of 347 GOPS/s. PipeLayer shows an average training throughput of 2923 GOPS/s and an average inference throughput of 3102 GOPS/s. The proposed *LiteCON* performs with an average training throughput of 90853 GOPS/s and an average inference throughput of 98958 GOPS/s. The superior performance of *LiteCON* is due to the intelligent integration of ultra-fast memristors and high-speed photonic components such as MDs, photodiodes, and DWDM waveguides.

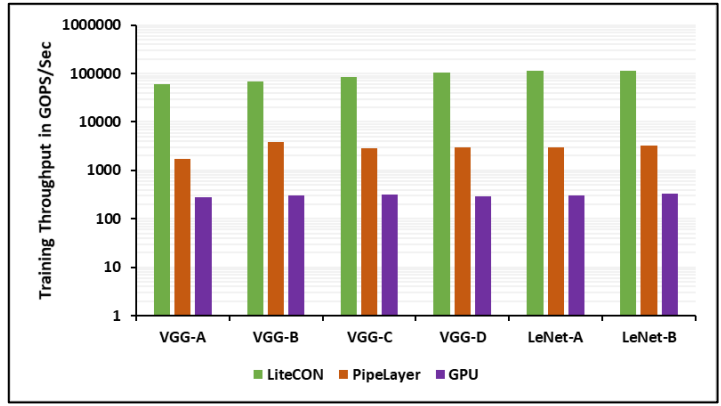

(a)

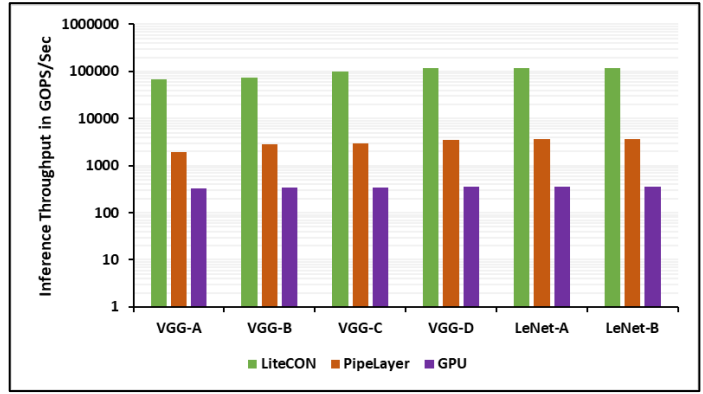

(b)

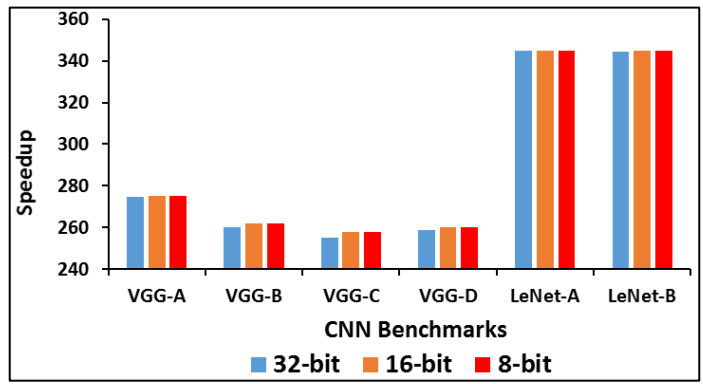

(c)

Figure 7: (a) Throughput comparison across accelerators during training; (b) Throughput comparison across accelerators during inference; (c) Speedup of *LiteCON* compared to GPU w.r.t. weight resolution.

The overall throughput of PipeLayer is affected by inter-layer data conversion with relatively slow ADCs. Also, PipeLayer spends most of its time in sequential weight updates during training. However, *LiteCON* has an inherent advantage due to its photonic parallel weight update mechanism. On average, *LiteCON* outperforms PipeLayer and GPU by 32× and 292× in terms of speedup, respectively. Finally, for the results presented in Fig. 7(a) and Fig. 7(b), the variance of throughput across benchmarks is 1650 with a standard deviation of 40.02 which is negligible considering the extreme scale throughput of *LiteCON*.

Fig.7c illustrates the effects of weight resolution on overall speedup of *LiteCON* compared to GPU. In general, weight resolution has negligible effect on the speedup of *LiteCON*. This is due to the fact that the data conversion (A-D or D-A) is done either at the beginning or at the end of the forward pass in *LiteCON*. Further, we see a slightly decreasing trend of speedup from VGG-A to VGG-D in Fig.7c. This is due to the increase in total number of convolution layers from VGG-A to VGG-D.

Fig.8 illustrates the computational efficiency per watt (CEPW) comparison of the proposed *LiteCON*, memristor crossbar based PipeLayer [6], and baseline GPU. For both training and inference, the CEPW trend is similar. Therefore, we show only one plot. PipeLayer uses memristor crossbars for the bulk of its arithmetic operations which has a CEPW of 120 GOPS/s/Watt/mm$^2$. However, the overall CEPW of PipeLayer comes down to 106 GOPS/s/Watt/mm$^2$ due to its extensive usage of data conversions. Also, ReLU and POOL are performed by a digital ALU in PipeLayer. This requires more memory to store intra-layer data for synchronizing with its pipeline mechanism. The superiority of *LiteCON* comes from the fact that it is a completely analog accelerator. Therefore, *LiteCON* does not involve inter-layer data conversions or storage for synchronization. AD and DA conversions are done either at the beginning or at the end of feature extraction in *LiteCON*. In addition to the compute efficient memristor, *LiteCON* also uses high speed OPAMP as ReLU. As shown in Fig.8, *LiteCON* has 5× and 60× higher computational efficiency compared to PipeLayer and GPU, respectively. The proposed *LiteCON* architecture shows a CEPW variance of 80.22 (standard deviation of 8.95 GOPS/s/Watt/mm$^2$) which is reasonable considering its high computational efficiency.

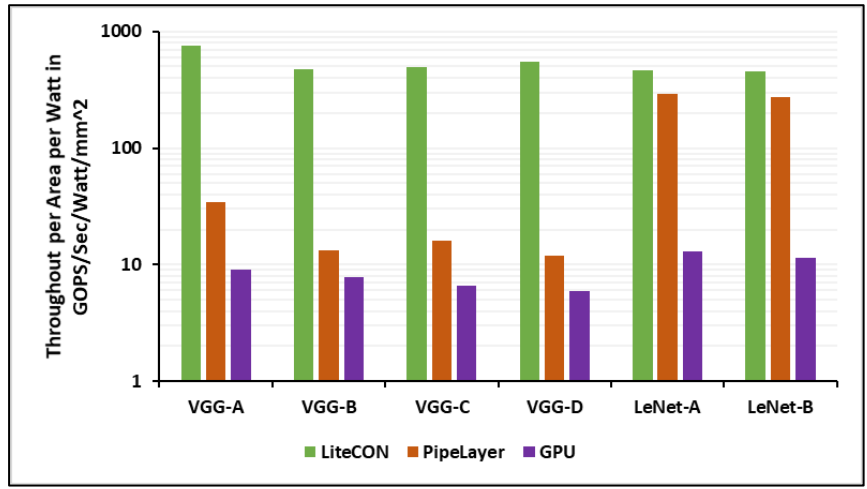

Figure 8: Computational efficiency per watt comparison of *LiteCON*, PipeLayer [6], and GPU [6].

## 6.3 ENERGY SAVINGS

We compare the energy efficiency of *LiteCON* with PipeLayer and GPU as depicted in Fig.9a and Fig.9b. For VGG-Net benchmarks, the average energy efficiency for PipeLayer is 31.3 GOPS/s/W and 33.2 GOPS/s/W during training and inference respectively. This is 1.5× and 1.7× higher than GPU based accelerator during training and inference respectively. For LeNet benchmarks, PipeLayer shows 21× and 22.7× higher energy efficiency compared to GPU. Unlike PipeLayer, *LiteCON* works uniformly across both VGG-Net and LeNet benchmarks with an energy efficiency of 1027.5 GOPS/s/W and 1096.5 GOPS/s/W during training and inference respectively. PipeLayer replicates its early feature extraction layers several times (close to 50K times) to maintain a balanced pipeline. This involves excessive use of high-power consuming data conversions. *LiteCON* uses passive optical components such as waveguides and microdisks, in addition to energy efficient components such as photodiodes and memristor. Also, *LiteCON* uses very few ADCs/DACs compared to PipeLayer. As shown in Fig.9a, for demanding benchmarks such as VGG-Net, we obtain 37× and 45× improvements in energy efficiency for *LiteCON* compared to PipeLayer and GPU, respectively. Overall, *LiteCON* outperforms PipeLayer and GPU for all benchmarks by 5× and 43×.

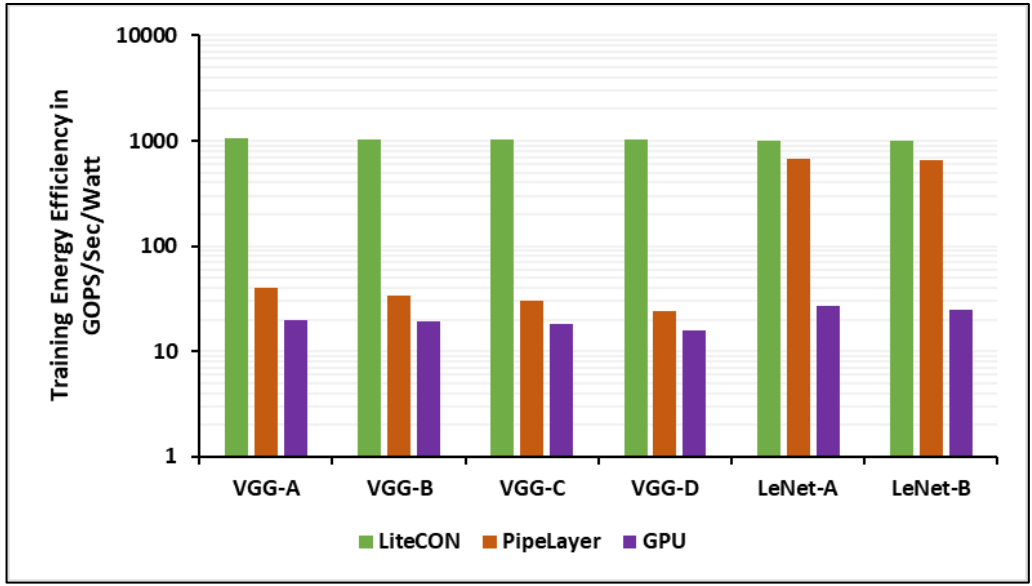

(a)

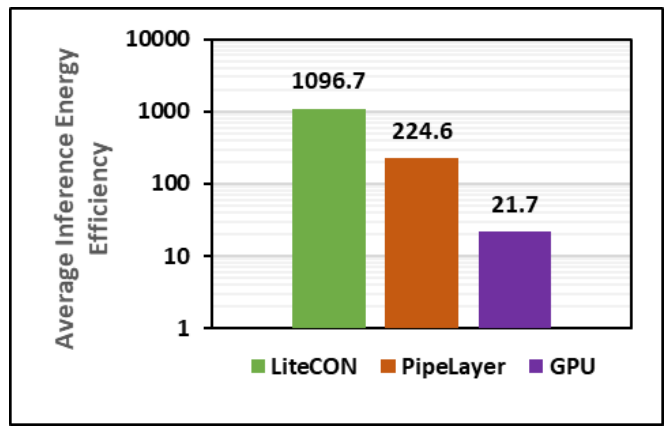

(b)

Figure 9: (a) Energy efficiency comparison across accelerators during training for VGG-Net and LeNet; (b) Average energy efficiency comparison across accelerators during inference.

## 6.4 COMPARISONS WITH LATEST PHOTONIC ACCELERATORS

Most of the photonic accelerators today deal only with inference. Therefore, we choose to compare the inference speedup with two promising photonic CNN accelerators [30] [31]. The comparison is illustrated in Table III.

TABLE III. COMPARISONS WITH LATEST PHOTONIC CNN ACCELERATORS [30][31]

| | [30] | [31] | *LiteCON* |
|---|---|---|---|
| Type of accelerator | Inference only | Inference only | Complete Accelerator (Both inference & training) |
| Fully analog, hybrid, digital | Fully analog | Hybrid (i.e., analog convolution; A/D conversion to DRAM; D/A conversion for analog convolution in the next layer) | Fully analog |
| Components | Convolution: Star coupler<br><br>Pooling: Star coupler<br><br>Activation: Opto-electric component | Convolution: Memristor banks | Convolution: Memristor<br><br>Pooling: Optical comparator<br><br>Activation: Optical amplifier |
| Compatibility | Only CNN | Only CNN | CNN (can be extended to DNN by making changes at design time) |
| Modulation involved | Both phase and amplitude | Only amplitude | Only amplitude |
| Type of activation | ReLU | ReLU | ReLU |
| Datasets | MNIST | MNIST | MNIST and ImageNet |
| Speedup w.r.t state-of-the-art GPU | Up to 65× | Up to 165× (for small scale MNIST datasets);<br><br>Up to 78× (for large-scale ImageNet dataset) | Up to 350× (for all datasets) |

## 6.5 PREDICTION ACCURACY

We performed a sensitivity analysis to investigate the impact of weight resolution on average prediction accuracy. Our design shows a prediction accuracy of 98% (i.e., slightly lower than state-of-the-art GPU accuracy of 99.3% and PipeLayer accuracy of 98.8%) for an 8-bit weight resolution. We choose this weight resolution because we use 8-bit DAC/ADC in our design. The prediction accuracy can be enhanced further by adopting an AD/DA mechanism with higher resolution. We choose not to at present to be on the conservative side from a CAD design standpoint. We take into account other sensitivity analysis such as effects of noises, propagation losses, photonic intrinsic losses, quantization error (in ADC/DAC), and quality factor of photonic components on prediction accuracy. The major factor among them is propagation loss that happens over the course of light signal traversal from the source to its destination. Fig. 10 shows the impact of propagation loss on accuracy.

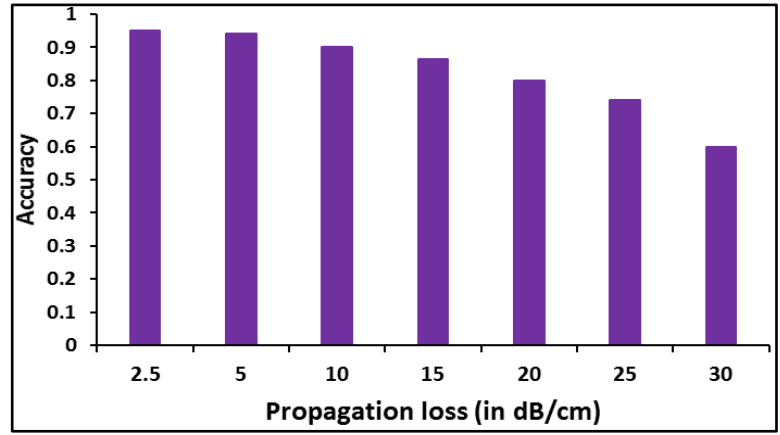

Fig. 10: Impact of propagation loss on prediction accuracy

For a 16-bit AD/DA resolution, *LiteCON* achieves 99.2% prediction accuracy for VGG and LeNet at the cost of 9% reduction in energy savings or energy efficiency. Anyway, the compromised energy efficiency with 16-bit resolution is still higher than the state-of-the-art PipeLayer and GPU (3.5× and 33× respectively, on average across benchmarks).

For a fair comparison, we brought down PipeLayer to 98% by considering 4-bit AD/DA resolution. This would enhance its energy efficiency by 10%, i.e., it increases from 273 GOPS/Sec/Watt (average) to 300 GOPS/Sec/Watt. The energy efficiency of GPU is not affected by changing the resolution as it is a completely digital system. So, the 300 GOPS/Sec/Watt is still less than *LiteCON*'s energy efficiency of 1132.85 GOPS/Sec/Watt (average).

Another factor that accounts for prediction accuracy of *LiteCON* is the finesse of the microdisk (MD) used in the system. Finesse determines the quality and operational accuracy of a microdisk; it depends on the intrinsic losses in an MD. Fig. 11 shows the impact of intrinsic losses (in dB/cm) on the finesse of MD of various sizes. The intrinsic loss of an MD depends upon the materials used. We assume an intrinsic loss of 2.5 dB/Cm in our design.

**Effects of component noise/error on accuracy**: the error/noise encountered by individual components play a role in determining the overall prediction error (PER). (1) Each memristor can have 1000 quantized states. The quantization error encountered due to limited number of memristor states contributes up to 1.2% of Prediction Error (PER); (2) The signal-to-noise ratio of microdisks used in *LiteCON* is 10 dB, which is adapted from [28]. The MD's contribution to the overall PER is 2.35%; (3) Each OPAMP in *LiteCON* has an SNR of 30 dB [29]. This accounts for a PER of 0.85%; and (4) the memristor-photonic interface is noisy. The signals from memristors going to modulators encounter a noise with an SNR of 25 dB which leads to a PER of 1.45%. We obtained these numbers through detailed optoelectronic synthesis using the IPKISS tool.

Please note that silicon photonic technology keeps evolving at a fast rate. With future improvement in microdisk finesse, intrinsic losses, and propagation loss, we can see accuracy close to state-of-the-art GPU accelerators (99.7% and beyond).

**Further improvement in accuracy by Incremental training**: Incremental training is a proven approach to enhance accuracy further and reduce training time. We performed incremental learning with *LiteCON* based on [27]. With such an approach, *LiteCON*'s accuracy increases from 98% to 98.7%. One major factor in the case of incremental training is storing previously-learned model or parameters in memory to be used in the next learning phase. To perform incremental training, we needed to include additional SRAM memory of 64KB.

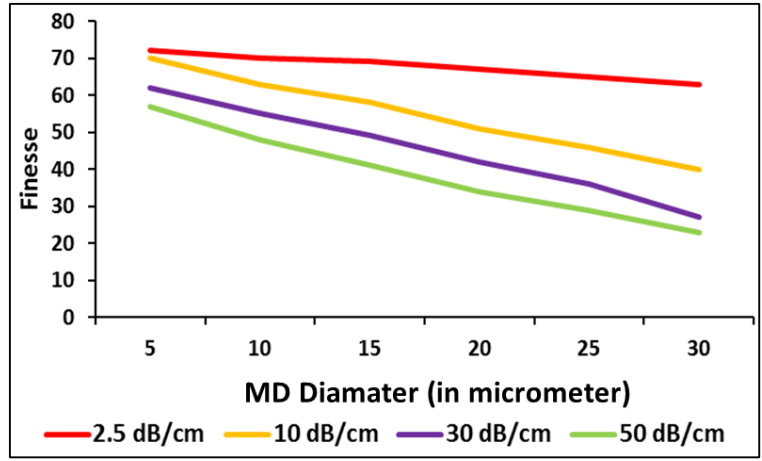

Fig. 11: Impact of intrinsic losses on the Finesse of a microdisk

## 6.6 DISCUSSION1: *LITECON* WITH COMPLEX MODELS

Now a days, more complex deep learning models are emerging, such as GoogleNet [24], Transformer [25], and BERT [26]. Architecture and characteristics wise they look extremely complex with 150+ hidden layers and millions of parameters. However, at the core of their functionalities, all of them comprise a softmax, an activation function, a fully connected layer, and a masking unit. *LiteCON* contains fundamental photonic components (ref: Section 4) to emulate these functionalities. For VGG and LeNet, we consider a ReLU activation; however, the activation circuit in *LiteCON* can be configured at design time to perform other neural activations as required by today's more complex deep learning models. One challenge that *LiteCON* would face while executing these large models is to perform multiple cycles of training without a long wait-time. That can be avoided by considering a multi-core *LiteCON* architecture connected by an optical on-chip network.

## 6.7 DISCUSSION2: EFFECTS OF MEMRISTOR AGING ON *LITECON*

Memristor plays a major role in *LiteCON*, i.e., to transfer analog data to the photonic realm in a pipelined fashion. This ensures the exemplary speedup of *LiteCON*. However, like any electrical device memristor also has many non-linear characteristics and is prone to degrade with aging. In *LiteCON*, we incorporated how an important non-idealities factor - aging, affects the performance of a memristor device. Being a non-reversible and inevitable process, it challenges the reliability of a memristor crossbar. We modeled an aging function to consider the effect of aging in a memristive device. We introduce a novel system-level aging model for memristor crossbars. Such a model can be integrated to any memristor CAD tool to investigate its performance accurately. In addition, we deploy an aging-aware memristor training scheme called skewed weight training. The proposed scheme incorporates age of each memristor cells to adjust their conductance matrix and current values dynamically thereby maintaining accuracy and energy efficiency. This is a first of its kind. Experiments with a standard CAD tool demonstrate 25% increase in the lifetime of a memristor crossbar by incorporating. The details of this work have been omitted due to brevity.

## 7. CONCLUSIONS

This paper demonstrates a fully analog CNN accelerator called *LiteCON* that optimally integrates low-area, ultra-fast, and energy-efficient photonic components such microdisks, waveguides, photodiodes, and splitter. *LiteCON* comprises a completely analog photonic backpropagation architecture. Further, the proposed architecture deploys (i) a scalable photonic convolution design based on microdisks in each CNN layer to emulate a range of sample CNN models; (ii) pipelined dataflow approach for high throughput. Compared to PipeLayer [6] and GPU, *LiteCON* architecture shows higher computational and energy efficiency due to the use of energy efficient microdisks and high-speed memristor crossbars, and also due to its use of a fully analog feature extraction method. We demonstrated that the proposed design has the potential to achieve up to 30×, 34×, and 4.5× improvements during training, and up to 34×, 40×, and 5.5× during inference, in throughput, energy efficiency, and computational efficiency per watt respectively, compared to the state-of-the-art with little reduction in accuracy. Our future work will address how *LiteCON* can be modeled for broader applicability such as other types of deep learning models, e.g., deep neural networks (DNNs).